\documentclass{article} 
\usepackage{iclr2021_conference,times}

\usepackage{amsmath,amsfonts,bm}

\def\eqref#1{equation~\ref{#1}}

\def\1{\bm{1}}

\DeclareMathAlphabet{\mathsfit}{\encodingdefault}{\sfdefault}{m}{sl}
\SetMathAlphabet{\mathsfit}{bold}{\encodingdefault}{\sfdefault}{bx}{n}

\usepackage{hyperref}
\usepackage{url}
\usepackage{mathtools}
\usepackage{bm}
\usepackage{esvect}
\usepackage{graphicx}
\usepackage{algorithm}
\usepackage{algorithmic}
\usepackage{lipsum}
\usepackage{graphics}

\title{FedPandemic: A Cross-Device Federated Learning Approach Towards Elementary Prognosis of Diseases During a Pandemic}

\author{Aman Priyanshu \\
Information Technology\\
Manipal Institute of Technology \\
\texttt{aman.priyanshu@learner.manipal.edu} \\
\And
Rakshit Naidu\thanks{School of Computer Science, Carnegie Mellon University \newline Correspondence to: $<$rnemakal@andrew.cmu.edu$>$} \\
Computer Science and Engineering\\
Manipal Institute of Technology \\
}

\iclrfinalcopy 
\begin{document}

\maketitle

\begin{abstract}
The amount of data, manpower and capital required to understand, evaluate and agree on a group of symptoms for the elementary prognosis of pandemic diseases is enormous. In this paper, we present FedPandemic, a novel noise implementation algorithm integrated with cross-device Federated learning for Elementary symptom prognosis during a pandemic, taking COVID-19 as a  case study. Our results display consistency and enhance robustness in recovering the common symptoms displayed by the disease, paving a faster and cheaper path towards symptom retrieval while also preserving the privacy of patients’ symptoms via Federated learning.
\end{abstract}

\section{Introduction}
Symptom prognosis and analysis are important tools of pandemic management, as medical conditions of the population could be gauged with these tools. However, appropriate symptoms and their exact effects were reported after mass collection and analysis during COVID-19 (\cite{Ghosh2020}, \cite{PandemicPreparednessreview}). This not only consumed time but also required an immense amount of manual effort to anonymize the continuously-growing large corpus of client data. In this paper, we propose \textit{FedPandemic}, a novel approach towards the elementary prognosis of diseases during a pandemic by cross-device Federated learning. We present a novel tool towards prominent symptom detection while retaining client privacy during an outbreak. This encourages collaborative efforts between the general public, smaller healthcare clinics/facilities, Non-Governmental Organizations (NGOs), hospitals and large network medical institutions. Federated learning (\cite{McMahanFedAvg16}, \cite{FLatScale19}) enables one to send models to where the data resides, rather than sending the data to the cloud thereby respecting the privacy of the users. Federated learning empowers distributed learning by gaining generalized insights over the active client space on decentralized data over a large number of rounds. \textit{FedPandemic} employs Word Embeddings as feature extractors for a binary classification model, which is trained using the Federated Averaging (FedAvg) Algorithm (\cite{McMahanFedAvg16}). The classifier is aimed to contribute towards preliminary medical examinations and prominent symptoms retrieval in the early stages of an outbreak. The model is developed in a mutable fashion to allow implementations of Secure Aggregation (\cite{SecAgg17}) or Differential Privacy (\cite{FLwithDP20}) for additional privacy use-cases.

\textit{FedPandemic} is trained based on the statistics of symptoms as reported by Statista's collection of COVID-19 symptoms in Kenya (\cite{KenyaStatista}), Germany (\cite{GermanyStatista}), Italy (\cite{ItalyStatista}), United States (\cite{USStatista}) and China (\cite{ChinaStatista}). The model employs and simulates different target clients with variable data sizes for learning. The implementation requires low computational prowess while still retaining high performance 
and client privacy making \textit{FedPandemic} a potentially strong tool towards future symptom detection during an outbreak.


We summarize five major problems presented in current symptom prognosis tools: (1) Time Consumption in centralized aggregation by a single institution. (2) Data Security of clients participating in such statistics. (3) Manual and Logistic Costs for data anonymization of COVID-19 data to protect client privacy. (4) Logistic, Manual and Infrastructural costs for over heading such a project. (5) Local Bias induced by smaller aggregators and analyzers.

Prominent symptom detection is an integral part of pandemic management and control. If these symptoms are detected and retrieved at the earliest, the process of elementary prognosis will be facilitated faster. This may allow different governments to prevent the spread of such diseases. However, current technologies, require a large network of people maintaining and analyzing this data, which is quite expensive. With \textit{FedPandemic}, we hope to overcome this problem using Federated learning to provide client privacy and low-cost maintenance based learning. 

\section{Proposed Methodology}
We employ Federated learning in a Cross-Device system, as this enables general public to contribute individually. The Federated Averaging algorithm (\cite{McMahanFedAvg16}) is used for generalizing the aggregated model. We utilize word embeddings for feature extraction on local devices which allows us to use State-Of-The-Art and 
also computationally resourceful encoders such as GloVe (\cite{Glove2014}) and Word2Vec embeddings. Word Embeddings produce a vector of fixed length as extracted features. This output is then fed into a client model, which is trained for a number of epochs $E$ and then the weights  of the updated model $w^{i}$ are returned to a centralised server. In this paper, we run multiple simulations on different contributors using GloVe (refer Table~\ref{tab:Table1}). 

\begin{figure}[h]
\centering
\includegraphics[width=14cm, keepaspectratio]{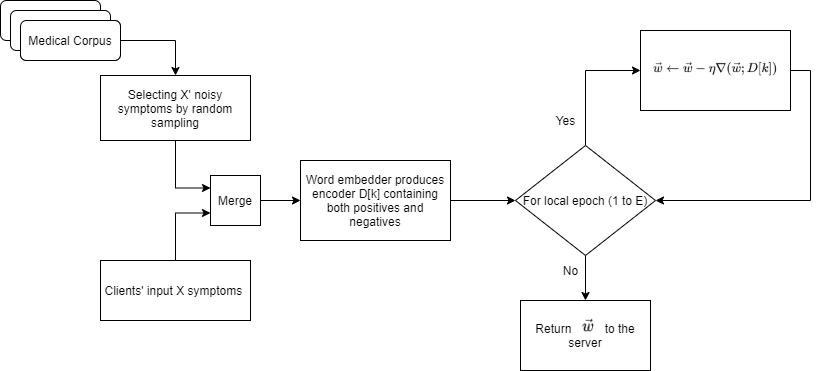}
\caption{Proposed Methodology of \textit{FedPandemic} for ClientUpdate in cross-device learning of prominent symptoms retrieval. The steps followed at the server are shown in Figure~\ref{fig:server_update}.}
\label{fig:Figure1}
\end{figure}

A common word encoder is decided for implementation and a lightweight classifier is designed keeping in mind the embedder selected. This allows us to develop a model, while at the same time keeping computational costs low. The selected embedder (here, GloVe) and model architecture are declared for training and aggregation. However, only training on client symptoms would make the classifier biased. Hence, we randomly sample symptoms from a given medical corpus, which are then learnt as negative samples by the models (refer Figure~\ref{fig:Figure1}).

The proposed methodology allows us to keep a pseudo data balance, thereby making our models robust to bias and underfitting. We believe that we propose the first implementation for symptom aggregation on a large-scale application that entertains both client-privacy as well as distributed learning. The procedure allows us to overcome some important issues of symptom analysis: (1) Manual aggregation of data from multiple healthcare centres is not required. (2) Common Symptoms that would be easily identified by the public, such as, high temperatures, fevers, cough and cold, would also be treated with prominence, giving the general public a better chance of discerning the infected. (3) Retains client privacy; evading efforts required for data anonymization. (4) Word Embeddings also allow semantically similar symptoms to be treated with prominence. This may aid researchers to study additional symptoms that the affected might be exhibiting.

\section{Experiments}
The experiments were conducted on a single system, running multiple instances of client models. The system consisted of 8GB RAM and a GeForce GTX 1650, 4GB GPU. We leverage the PyTorch framework for our experiments and the base algorithm used for Federated Learning was FedAvg (\cite{McMahanFedAvg16}). The classifier used in our experiments consisted of (50, 32, 16, 8, 1) neurons from the top layer to the bottom layer. For our experiments, the learning rate and batch size were chosen as 0.001 and 32 respectively along with the Adam optimizer. Our approach involved
four simulations represented by different aggregation steps, which can be employed by local authorities. Our presentation takes statistical numbers from data; as published on Statista. We choose GloVe (\cite{Glove2014}) as our encoder in our experiments because its embeddings are light-weight and easy-to-use in a Federated learning environment when compared to embeddings from other State-of-the-art encoders like BERT (\cite{BERT18}) and ELMO (\cite{ELMO18}).

In this work, we present four variants of simulations (see Table~\ref{tab:Table1}):

\begin{itemize}
    \item \textbf{Simulation I:} \underline{Large Medical Institutes (Baseline)}
    
    This simulation aims towards reproducing aggregation by large medical institutes. In this simulation, we distribute the entire corpus, into 20 institutions or clients and train a federated model. Each institution has been given an equal number of sample cases (60,000 samples). This simulation is definite as large medical institutes will already have enough data to ensure that they can select which symptoms are prominent.
    
    \item \textbf{Simulation II:} \underline{Medium Ranged Medical Institutes, like Hospitals, NGOs, etc.}
    
    This simulation offers to cluster and pick symptoms from a larger collaborating group. However, even this group is large enough to accurately classify prominent symptoms. In this case, the data is not equally distributed and ranges between 10,000 to 20,000 samples.
    
    \item \textbf{Simulation III:} \underline{Small Ranged Medical Institutes, like clinics and health care centres}
    
    This simulation is the most practical one, as these institutes may be able to actively collaborate for training such a model. Each client will have samples ranging from 500 to 2,000.
    
    \item \textbf{Simulation IV:} \underline{Individual/Family Contributions}
    
    This simulation is the toughest to learn and provides the most realistic sample which could be implemented for the preliminary search of symptoms. Each client contains samples between 2 and 12.
\end{itemize}

\begin{figure}[h]
\centering
\includegraphics[width=6cm]{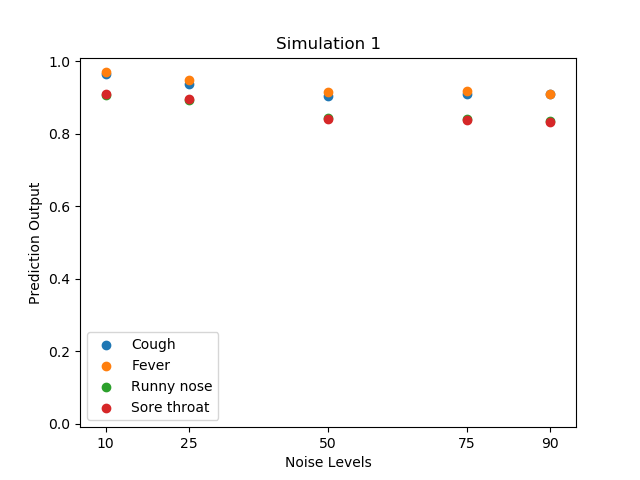}
\includegraphics[width=6cm]{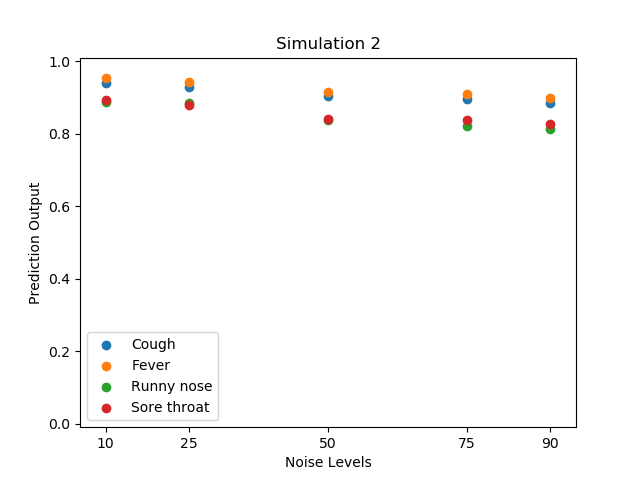}
\includegraphics[width=6cm]{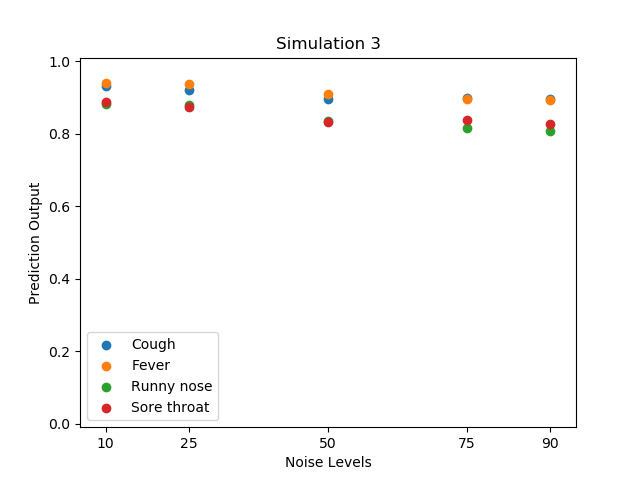}
\includegraphics[width=6cm]{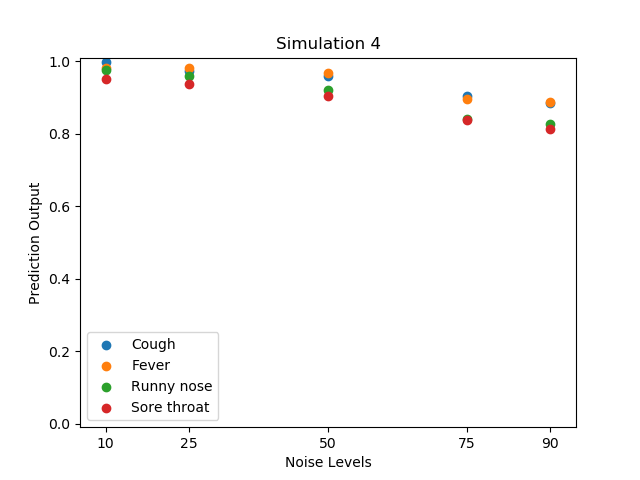}

\caption{Prediction output vs Noise levels for four of the most common symptoms across all countries sampling from the Normal distribution $N\sim{(0,1)}$.}
\label{fig:normal_sims}
\end{figure}

These simulations are pulled from the given distribution (refer Figure~\ref{fig:bargraphsymp}) and aim to replicate real-world usage of \textit{FedPandemic}. We provide experimental results on a few prominent symptoms with different noise levels against the prediction output (refer Figure~\ref{fig:normal_sims}). 

\begin{figure}[h]
\centering
\includegraphics[width=6cm]{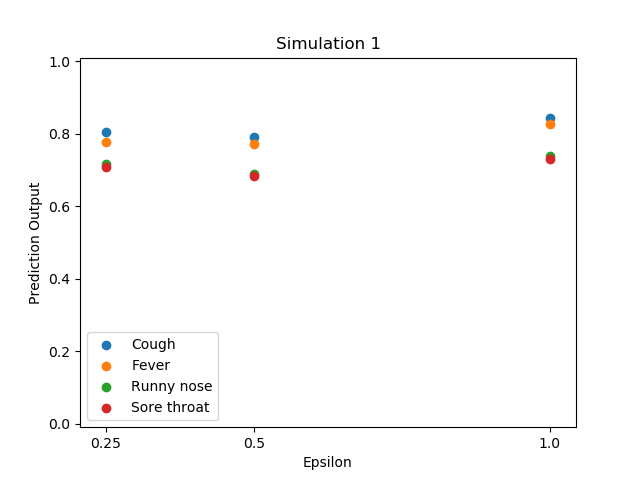}
\includegraphics[width=6cm]{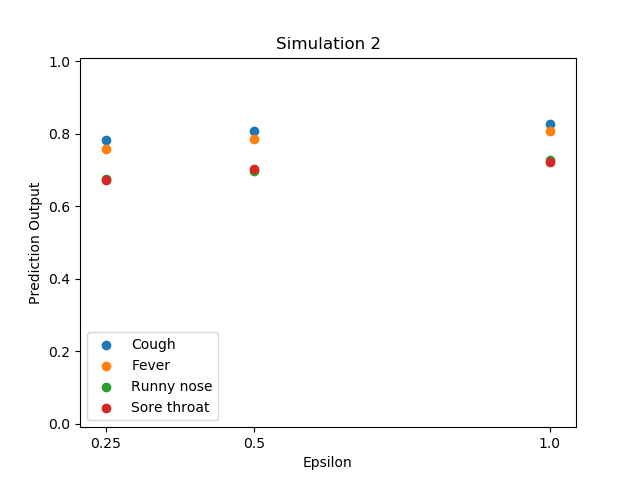}
\includegraphics[width=6cm]{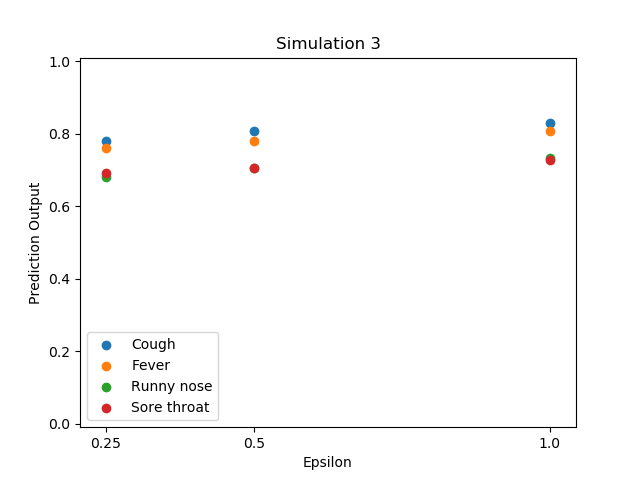}
\includegraphics[width=6cm]{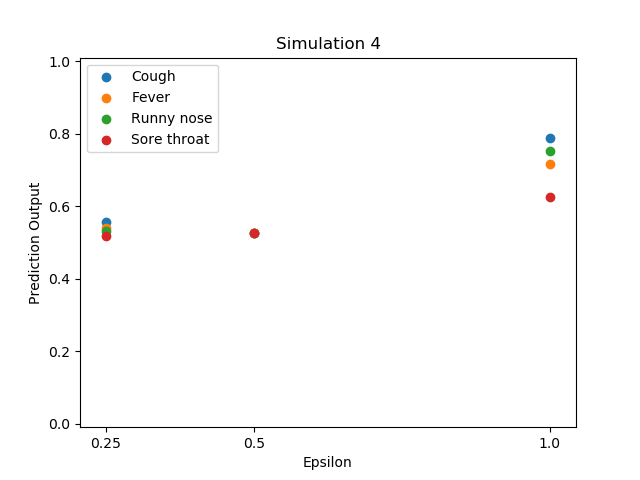}

\caption{Prediction output vs Epsilon values for four of the most common symptoms across all countries sampling from the Laplace distribution $L\sim{(0,1/\epsilon)}$.}
\label{fig:lap_sims}
\end{figure}

We experiment the random sampling step (refer Figure~\ref{fig:Figure1}) with Normal and Laplacian Distribution values. The Laplace mechanism (with a paramter of $\frac{1}{\epsilon}$) preserves $\epsilon$-Differential Privacy (\cite{AlgDiffPrivacy_DworkRoth14}). We vary different values of $\epsilon$ (taking 50\% as the noise level as standard across all the simulations) to observe how our algorithm plays into Differential Privacy guarantees (as shown in Figure~\ref{fig:lap_sims}).

We also display experiment results for different noise levels for target symptoms shown by greater than and lesser than 10\% of the Survey Population (refer Figure~\ref{fig:0greater10}). 

\section{Conclusion and Future work}
In this paper, we showcase a novel approach using Federated learning towards Elementary Symptom Prognosis in order to preserve client privacy and improve faster response times during a pandemic. Our experiments include various noise levels and the accuracy levels drop consistently as the noise values are increased (refer Table~\ref{tab:Table2}). Simulation IV displays highest output predictions as we evaluate over a large client space which signifies more personalized models (see Figure~\ref{fig:normal_sims}). We see that the Laplacian variant of our algorithm provides $\epsilon$-DP (as $\epsilon$ increases, lower amount of noise is added which intuitively means higher utility and higher accuracy) in Figure~\ref{fig:lap_sims}. We believe that either Simulation II and III could make best use of $FedPandemic$ (given their size range and number of clients)  with the 50\% noise level readings by best replicating real-world situations. We hope to improve our method by making it robust to malicious attacks and Byzantine failures. We wish to improve the training model by incorporating data from other countries as well. 

\bibliography{fedpandemic_ref}
\bibliographystyle{iclr2021_conference}

\appendix
\section{Appendix}
\subsection{Noise Implementation}

We present multiple simulations with different levels of noise simulated. In our experiments, we introduce five different noise levels (We include 10\% and 90\% readings as shown in Figure~\ref{fig:noisevspred} to display results with extreme noise levels) :
\begin{enumerate}
    \item \underline{\textbf{10}}: Almost perfect simulation, with minimal amount of noise. The experiment can be thought of as a setting where only individuals infected with the novel coronavirus or the pandemic in consideration. There is little to no influence of any other symptoms the clients may have been facing during learning.
    
    \item \underline{\textbf{25}}: Close to ideal simulation. Here the noise levels are increased by 25\%. That is, an individual may report symptoms other than that from the disease in consideration with a 25\% probability. This setting is closer to reality, as the general public would not know whether the symptoms they feel are relevant to the pandemic or not.

    \item \underline{\textbf{50}}: Here noise levels have been set to 50\%. That is, an individual may report symptoms other than that from the disease in consideration with a 50\% probability. A step closer to reality and probably the closest, as most citizens are still healthy or if suffering would recognize new symptoms easily. However, noise would still be generated due to the large sample space.

    \item \underline{\textbf{75}}: Noise levels are set to 75\%. In this case the citizens have a higher chance of entering symptoms that are not related to the pandemic in question, however, due to the frequency presented in the total population, insignificant/not associated symptoms would be lost.

    \item \underline{\textbf{90}}: If a person does not put a symptom related to the virus, they will put another arbitrary symptom. Using 90\% noise level, we show that this is entirely based on the frequency analysis shown during a pandemic. The only reason our federated model is expected to converge and learn in such a case, is the sheer frequency of pandemic victims. Therefore, we specifically target our project towards pandemics like novel coronavirus.
\end{enumerate}


\begin{algorithm}[H]
  \caption{Noise Implementation}
  \label{alg:noise_algo}
\begin{algorithmic}[1]
\STATE {\bfseries Input:} 
\STATE Survey\_Population $\leftarrow$ \textit{People using the application}; \\

\STATE Prominent\_COVID19\_symptoms $\leftarrow$ [cs$_1$, cs$_2$, ..., cs$_n$]; \textit{the actual symptoms displayed as per the research conducted by Statista. Ex: [Fever, Cough, Headache]} \\

\STATE Symptom\_Probability $\leftarrow$ [ps$_1$, ps$_2$, ..., ps$_n$]; \textit{probability of prominent symptoms that was displayed by the research conducted by Statista. Ex: [0.37, 0.2, 0.1]} \\

\STATE Medical\_Corpus $\leftarrow$ [s$_1$, s$_2$, ..., s$_m$, cs$_1$, cs$_2$, ..., cs$_n$]; \textit{medical corpus present in GloVe which the public may identify. Ex: [Stomach Ache, Anemia, Red Eyes, ..., Fever, Cough, Headache, ..., ulcers, etc.]}

\STATE random.random() $\leftarrow$ \textit{random number sampled from the normal distribution N(0,1).}

{\bfseries Algorithm:}
\FOR{$i\in$ Survey\_Population}
\STATE Symptoms\_Displayed = []; \\

\FOR{$s\in$ Prominent\_COVID19\_Symptoms}

    \IF{random.random() $<$ Symptom\_Probability[$s$]} 
        \STATE Symptoms\_Displayed.append($s$); \\
    \ELSIF{random.random() $<$ Noise\_Level*}
        \STATE Symptoms\_Displayed.append(random\_symptom(Medical\_Corpus)); \\
    \ENDIF

\ENDFOR \\
\ENDFOR \\

\end{algorithmic}
\end{algorithm}

\let\thefootnote\relax\footnote{*For the Laplacian mechanism, we employ `random.laplace(0, 1/$\epsilon$) $>$ Noise\_Level` in order to satisfy $\epsilon$-DP properties.}

\begin{figure}[h]
\centering
\includegraphics[width=15cm, keepaspectratio]{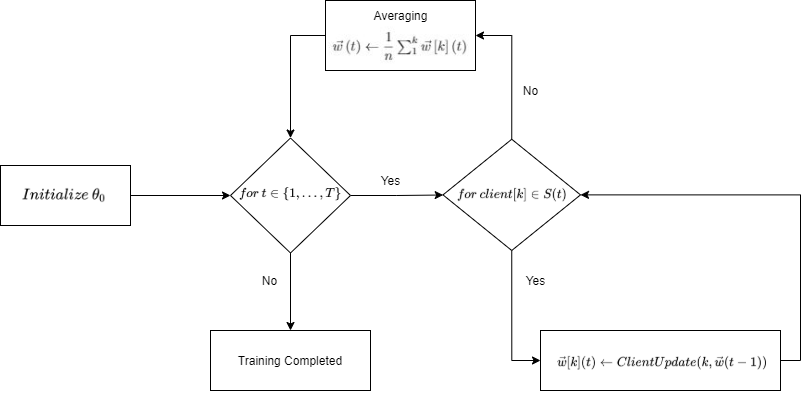}

\caption{Function of the Central Server for running \textit{FedPandemic}.}
\label{fig:server_update}
\end{figure}

\begin{table}[h]
\centering
\begin{tabular}{|l|l|l|l|l|l|}
\hline
\textbf{Simulations} & Size Range     & No. of clients & Local epochs & Global epochs  \\ \hline
I                    & (60000, 60000) & 20             & 5            & 5    \\ \hline
II                   & (10000, 20000) & 80             & 5            & 5    \\ \hline
III                  & (500, 2000)    & 900            & 5            & 5    \\ \hline
IV                   & (2, 12)        & 100000         & 5            & 5\\ \hline
\end{tabular}
\caption{Simulations characteristics}
\label{tab:Table1}
\end{table}

\begin{table}[h]
\centering
\begin{tabular}{ll}
\hline
\multicolumn{1}{|l|}{\textbf{Noise Levels}} & \multicolumn{1}{l|}{Accuracy} \\ \hline
\multicolumn{1}{|l|}{0\%}                   & \multicolumn{1}{l|}{1.0}      \\ \hline
\multicolumn{1}{|l|}{25\%`}                 & \multicolumn{1}{l|}{0.75}     \\ \hline
\multicolumn{1}{|l|}{50\%}                  & \multicolumn{1}{l|}{0.75}     \\ \hline
\multicolumn{1}{|l|}{75\%}                                        & \multicolumn{1}{l|}{0.6875}         \\
\hline
\multicolumn{1}{|l|}{100\%}                                       & \multicolumn{1}{l|}{0.5625} \\          
\hline
\end{tabular}
\caption{Noise Levels vs Accuracy after four global epochs}
\label{tab:Table2}
\end{table}

\begin{table}[h]
\centering
\begin{tabular}{|l|l|l|l|l|l|}
\hline
Country                & USA    & China & Germany & Italy & Kenya \\ \hline
Total                  & 373883 & 55924 & 747900  & 34142 & 14616 \\ \hline
Fever                  & 161330 & 49157 & 209412  & 25606 & 13161 \\ \hline
Cough                  & 186941 & 37860 & 299160  & 12973 & 10815 \\ \hline
Shortness of breath    & 104687 & 10401 & 0       & 0     & 0     \\ \hline
Myalgia                & 134784 & 0     & 0       & 0     & 0     \\ \hline
Runny nose             & 22619  & 18678 & 194454  & 0     & 9025  \\ \hline
Sore throat            & 74402  & 7773  & 157059  & 0     & 4194  \\ \hline
Headache               & 127867 & 7605  & 0       & 0     & 8038  \\ \hline
Nausea/Vomiting        & 42435  & 2796  & 0       & 0     & 0     \\ \hline
Abdominal pain         & 28228  & 0     & 0       & 0     & 0     \\ \hline
Diarrhea               & 71037  & 2069  & 0       & 2048  & 0     \\ \hline
Loss of smell or taste & 30845  & 0     & 157059  & 0     & 0     \\ \hline
Fatigue                & 0      & 21307 & 0       & 0     & 0     \\ \hline
Muscle Pain            & 0      & 8276  & 0       & 0     & 0     \\ \hline
Chills                 & 0      & 6375  & 0       & 0     & 0     \\ \hline
Nasal Congestion       & 0      & 2684  & 0       & 24923 & 10231 \\ \hline
Pneumonia              & 0      & 0     & 7479    & 0     & 0     \\ \hline
\end{tabular}
\caption{\label{tab:Table2}This table consists of absolute values of the people showing these symptoms according to country. We extract the base probability distribution on which the simulations are performed.}
\end{table}

\subsection{Experimental Data}

\begin{table}[h]
\centering
\begin{tabular}{|l|l|l|l|}
\hline
\multicolumn{1}{|c|}{Country}        & \multicolumn{1}{c|}{USA} & \multicolumn{1}{c|}{Country} & \multicolumn{1}{c|}{Kenya} \\ \hline
Total                                & 3,73,883                 & Total                        & 14,616                     \\ \hline
Fever, cough, or shortness of breath & 69.8\%                   & Fever                        & 90.05\%                    \\ \hline
Fever                                & 43.15\%                  & Dry cough                    & 74\%                       \\ \hline
Cough                                & 50\%                     & Difficulty in breathing      & 70\%                       \\ \hline
Shortness of breath                  & 28\%                     & Sneezing                     & 61.75\%                    \\ \hline
\end{tabular}
\end{table}

\begin{table}[h]
\centering
\begin{tabular}{|l|l|l|l|l|l|}
\hline
\multicolumn{1}{|c|}{Country}            & \multicolumn{1}{c|}{China}  & \multicolumn{1}{c|}{Country}     & \multicolumn{1}{c|}{Germany}  & \multicolumn{1}{c|}{Country}  & \multicolumn{1}{c|}{Italy} \\\hline
Total              & 55924  & Total       & 7,47,900 & Total    & 34142 \\\hline
Fever              & 87.9\% & Cough       & 40\%     & Fever    & 75\%  \\\hline
Dry cough          & 67.7\% & Fever       & 28\%     & Dyspnoea & 73\%  \\\hline
Fatigue            & 38.1\% & Runny nose  & 26\%     & Cough    & 38\%  \\\hline
Sputum production & 33.4\% & Sore throat & 21\%     & Diarrhea & 6\%  \\\hline
\end{tabular}
\caption{\label{tab:Table3}These are the 4 most prominent symptoms as well as the total number of participants from every country included in our dataset.}
\end{table}

\begin{figure}[h]
\includegraphics[width=15cm]{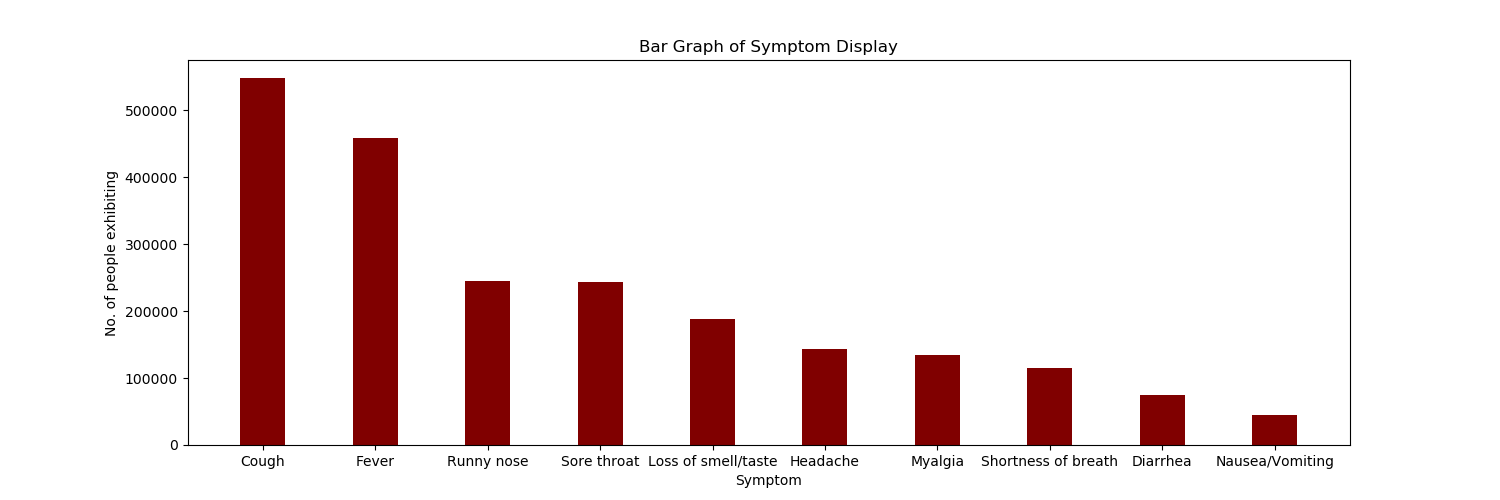}
\caption{This figure is bar graph representation, of the total populations displaying said symptoms. In retrospect, these values represent the general display such symptoms as the total number of participates makes up 1.2 Billion people. The perturbations are generated from the same distribution in order to approach realistic sample spaces.}
\label{fig:bargraphsymp}
\end{figure}

For the work presented in this paper, we employ simulations for data collection and cross-device setup. As there is no similar objective or dataset, the paper has taken the liberty to implement these simulations based on real-world data recordings. We take the data collected by Statista for COVID-19 symptoms in different countries. The data presented is a statistical representation of the \% of people exhibiting a certain symptom. The total number of samples taken from the entire corpus makes up 1,226,465 people's data. The countries whose data has been used for simulation are:

\begin{itemize}
    \item Kenya
    \item Germany
    \item Italy
    \item United States
    \item China
\end{itemize}

We present the statistics of each of these countries in Table~\ref{tab:Table1}, and the aggregate statistics we pulled in Table~\ref{tab:Table2}. The data presented in Table~\ref{tab:Table2} has been visualized (as a bar graph) in Figure~\ref{fig:bargraphsymp}.

\begin{figure}[h]
\centering
\includegraphics[width=13cm,keepaspectratio]{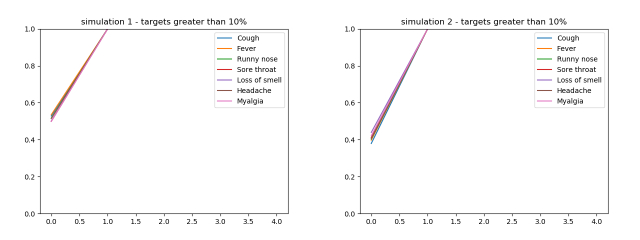}
\includegraphics[width=13cm,keepaspectratio]{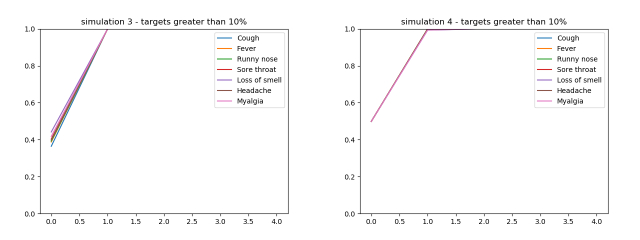}
\label{fig:0greater10}
\caption{Simulations for 0\% noise and target Symptoms exhibited by greater than 10\% of the Survey population. [X-Axis: Number of global epochs; Y-Axis: Accuracy]}

\end{figure}

\begin{figure}[h]
\centering
\includegraphics[width=13cm,keepaspectratio]{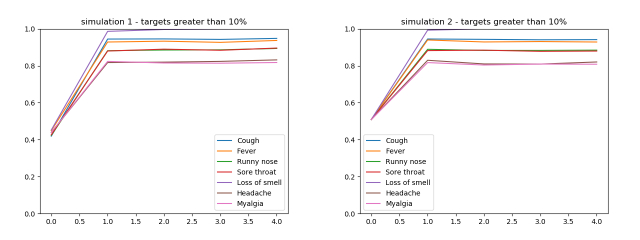}
\includegraphics[width=13cm,keepaspectratio]{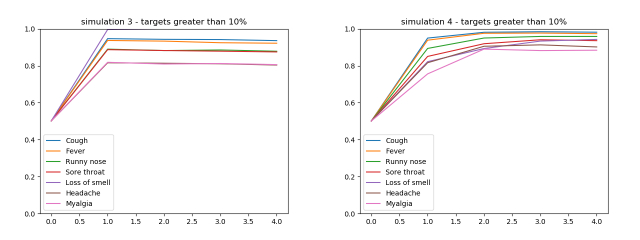}

\caption{Simulations for 25\% noise and target Symptoms exhibited by greater than 10\% of the Survey population. [X-Axis: Number of global epochs; Y-Axis: Accuracy]}

\end{figure}

\begin{figure}[h]
\centering
\includegraphics[width=13cm,keepaspectratio]{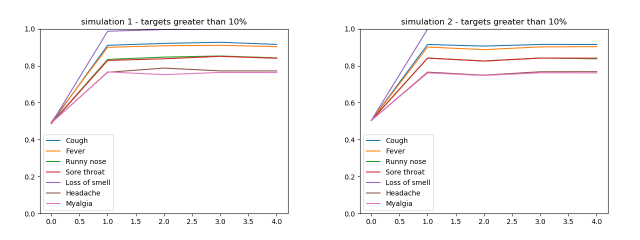}
\includegraphics[width=13cm,keepaspectratio]{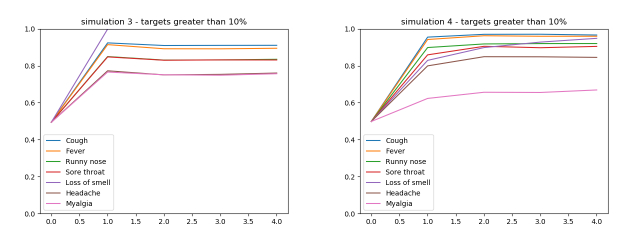}

\caption{Simulations for 50\% noise and target Symptoms exhibited by greater than 10\% of the Survey population. [X-Axis: Number of global epochs; Y-Axis: Accuracy]}

\end{figure}

\begin{figure}[h]
\centering
\includegraphics[width=13cm,keepaspectratio]{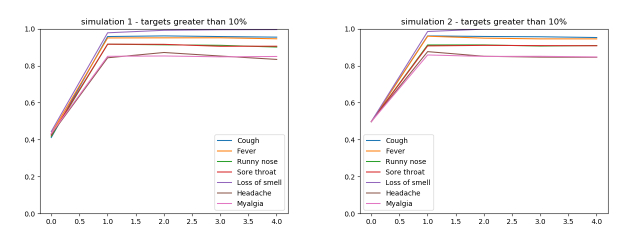}
\includegraphics[width=13cm,keepaspectratio]{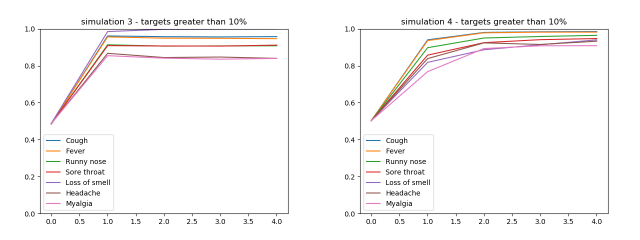}

\caption{Simulations for 75\% noise and target Symptoms exhibited by greater than 10\% of the Survey population. [X-Axis: Number of global epochs; Y-Axis: Accuracy]}

\end{figure}

\begin{figure}[h]
\centering
\includegraphics[width=13cm,keepaspectratio]{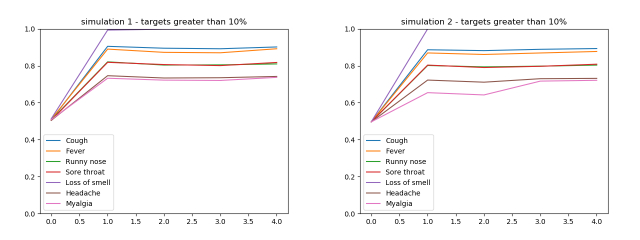}
\includegraphics[width=13cm,keepaspectratio]{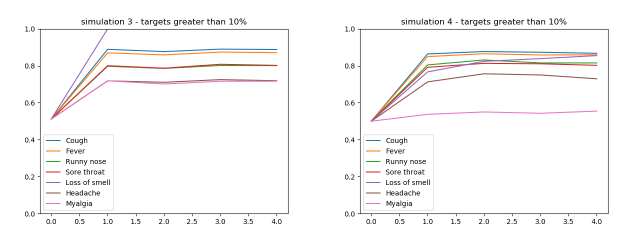}

\caption{Simulations for 100\% noise and target Symptoms exhibited by greater than 10\% of the Survey population. [X-Axis: Number of global epochs; Y-Axis: Accuracy]}

\end{figure}

\begin{figure}[h]
\centering
\includegraphics[width=13cm,keepaspectratio]{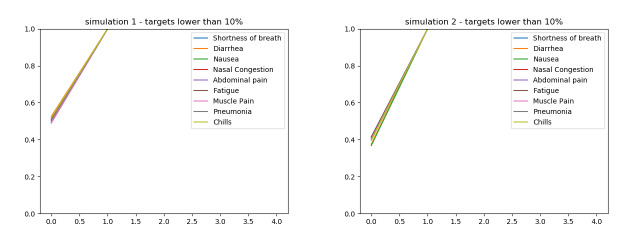}
\includegraphics[width=13cm,keepaspectratio]{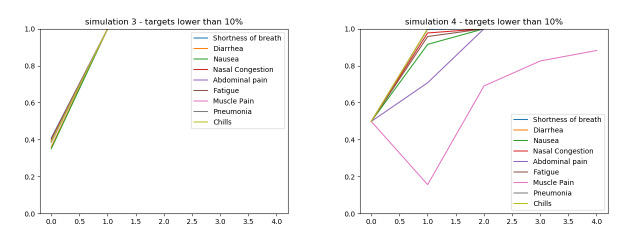}

\caption{Simulations for 0\% noise and target Symptoms exhibited by lesser than 10\% of the Survey population. [X-Axis: Number of global epochs; Y-Axis: Accuracy]}

\end{figure}

\begin{figure}[h]
\centering
\includegraphics[width=13cm,keepaspectratio]{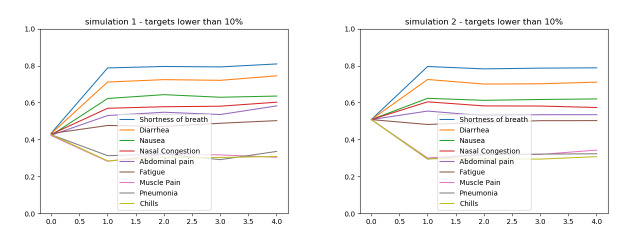}
\includegraphics[width=13cm,keepaspectratio]{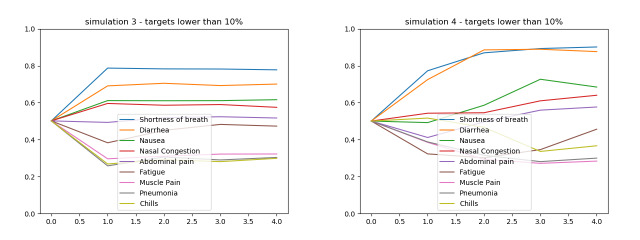}

\caption{Simulations for 25\% noise and target Symptoms exhibited by lesser than 10\% of the Survey population. [X-Axis: Number of global epochs; Y-Axis: Accuracy]}

\end{figure}

\begin{figure}[h]
\centering
\includegraphics[width=13cm,keepaspectratio]{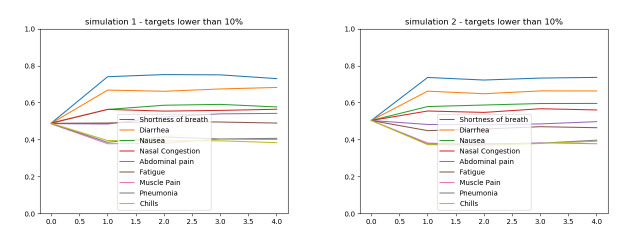}
\includegraphics[width=13cm,keepaspectratio]{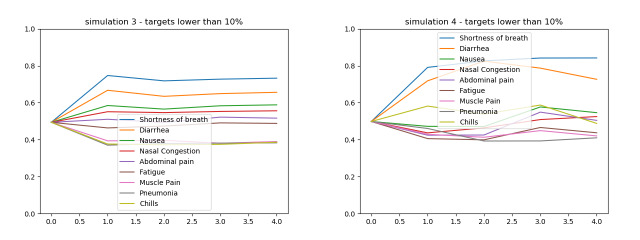}

\caption{Simulations for 50\% noise and target Symptoms exhibited by lesser than 10\% of the Survey population. [X-Axis: Number of global epochs; Y-Axis: Accuracy]}

\end{figure}

\begin{figure}[h]
\centering
\includegraphics[width=13cm,keepaspectratio]{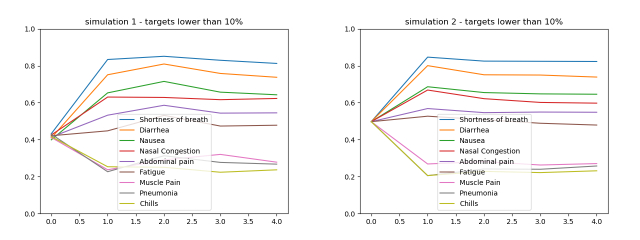}
\includegraphics[width=13cm,keepaspectratio]{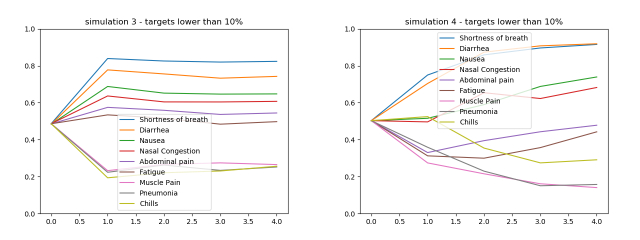}

\caption{Simulations for 75\% noise and target Symptoms exhibited by lesser than 10\% of the Survey population. [X-Axis: Number of global epochs; Y-Axis: Accuracy]}

\end{figure}

\begin{figure}[h]
\centering
\includegraphics[width=13cm,keepaspectratio]{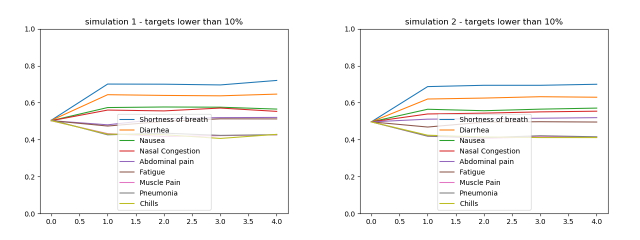}
\includegraphics[width=13cm,keepaspectratio]{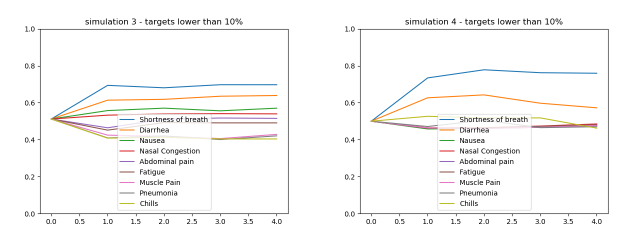}
\label{fig:100lesser10}
\caption{Simulations for 100\% noise and target Symptoms exhibited by lesser than 10\% of the Survey population. [X-Axis: Number of global epochs; Y-Axis: Accuracy]}

\end{figure}

\end{document}